\documentclass[reprint,superscriptaddress,groupedaddress,amsmath]{revtex4}  
\usepackage{graphicx}  
\usepackage{dcolumn}   
\usepackage{bm}        
\usepackage{amssymb}   
\usepackage{xcolor} 

\setcitestyle{round} 

\begin{document}

\title{\textbf{Origin of the butterfly magnetoresistance in a Dirac nodal-line system}}
\author{\textbf{Y. -C. Chiu}}
\affiliation{National High Magnetic Field Laboratory, Florida State University, Tallahassee-FL 32310, USA}
\affiliation{Department of Physics, Florida State University, Tallahassee-FL 32306, USA}
\author{\textbf{K. -W. Chen}}
\affiliation{National High Magnetic Field Laboratory, Florida State University, Tallahassee-FL 32310, USA}
\affiliation{Department of Physics, Florida State University, Tallahassee-FL 32306, USA}
\author{\textbf{R. Sch\"{o}nemann}}
\affiliation{National High Magnetic Field Laboratory, Florida State University, Tallahassee-FL 32310, USA}
\author{\textbf{V. L. Quito}}
\affiliation{National High Magnetic Field Laboratory, Florida State University, Tallahassee-FL 32310, USA}
\affiliation{Department of Physics, Florida State University, Tallahassee-FL 32306, USA}
\affiliation{Department of Physics and Astronomy, Iowa State University, Ames, Iowa 50011, USA}
\author{\textbf{S. Sur}}
\affiliation{National High Magnetic Field Laboratory, Florida State University, Tallahassee-FL 32310, USA}
\affiliation{Department of Physics, Florida State University, Tallahassee-FL 32306, USA}
\affiliation{Department of Physics \& Astronomy, Northwestern University, Evanston, IL 60208, USA}
\author{\textbf{Q. Zhou}}
\affiliation{National High Magnetic Field Laboratory, Florida State University, Tallahassee-FL 32310, USA}
\affiliation{Department of Physics, Florida State University, Tallahassee-FL 32306, USA}
\author{\textbf{D. Graf}}
\affiliation{National High Magnetic Field Laboratory, Florida State University, Tallahassee-FL 32310, USA}
\author{\textbf{E. Kampert}}
\affiliation{Dresden High Magnetic Field Laboratory (HLD-EMFL), Helmholtz-Zentrum Dresden-Rossendorf, 01328 Dresden, Germany}
\author{\textbf{T. F\"{o}rster}}
\affiliation{Dresden High Magnetic Field Laboratory (HLD-EMFL), Helmholtz-Zentrum Dresden-Rossendorf, 01328 Dresden, Germany}
\author{\textbf{K. Yang}}
\affiliation{National High Magnetic Field Laboratory, Florida State University, Tallahassee-FL 32310, USA}
\affiliation{Department of Physics, Florida State University, Tallahassee-FL 32306, USA}
\author{\textbf{G. T. McCandless}}
\affiliation{The University of Texas at Dallas, Department of Chemistry and Biochemistry, Richardson, TX 75080 USA}
\author{\textbf{Julia Y. Chan}}
\affiliation{The University of Texas at Dallas, Department of Chemistry and Biochemistry, Richardson, TX 75080 USA}
\author{\textbf{R. E. Baumbach}}
\affiliation{National High Magnetic Field Laboratory, Florida State University, Tallahassee-FL 32310, USA}
\affiliation{Department of Physics, Florida State University, Tallahassee-FL 32306, USA}
\author{\textbf{M. D. Johannes}}
\affiliation{Center for Computational Materials Science, Naval Research Laboratory, Washington, DC 20375, USA}
\author{\textbf{L. Balicas}}
\affiliation{National High Magnetic Field Laboratory, Florida State University, Tallahassee-FL 32310, USA}
\affiliation{Department of Physics, Florida State University, Tallahassee-FL 32306, USA}
\date{\today}

\begin{abstract}
\textbf{We report a study on the magnetotransport properties and on the Fermi surfaces (FS) of the ZrSi(Se,Te) semimetals. Density Functional Theory (DFT) calculations, in absence of spin orbit coupling  (SOC), reveal that both the Se and the Te compounds display Dirac nodal lines  (DNL) close to the Fermi level  $\varepsilon_F$ at symmorphic and non-symmorphic positions, respectively.  We find that the geometry of their FSs agrees well with DFT predictions.  ZrSiSe displays low residual resistivities, pronounced magnetoresistivity, high carrier mobilities, and a butterfly-like angle-dependent magnetoresistivity (AMR), although its DNL is not protected against gap opening. As in Cd$_3$As$_2$, its transport lifetime is found to be $10^2$ to $10^3$ times larger than its quantum one. ZrSiTe, which possesses a protected DNL, displays conventional transport properties. Our evaluation indicates that both compounds most likely are topologically trivial. Nearly angle-independent effective masses with strong angle dependent quantum lifetimes lead to the butterfly AMR in ZrSiSe}
\end{abstract}
\maketitle

\section{Introduction}
In the last decade, much attention has been devoted to the discovery of semimetallic systems whose electronic band structures display features protected by the interplay of symmetry and topology \cite{fu,review2,dirac_semi_metals,TP_transition,rappe_dirac_line_nodes}. Many of these compounds, such as the topological insulators \cite{fu,review2}, the Dirac \cite{Na3Bi_discovery,ong_Cd3As2} and the Weyl semimetals \cite{hasan_weyl1,hasan_TaAs,hasan_NbAs,Lu,type_II}, are characterized by strong spin-orbit coupling and robust gapless surface states resulting from their nontrivial band structure topology. As exemplified by graphene, topology related phenomena can also occur in materials characterized by a very small spin-orbit interaction \cite{magnetic}. Among these, Dirac nodal-line semimetals are compounds whose valence and conduction bands linearly touch at a collection of nodal points depicting a one-dimensional nodal line (NL) within the bulk Brillouin zone \cite{rappe_dirac_line_nodes}. The stability of these NLs require the absence of spin-orbit coupling (SOC) since its inclusion tends to gap them out due to the mixing of spin components. The existence of the NLs in the presence of SOC requires the protection from extra crystalline symmetries, specifically non-symmorphic ones. Proposals for Dirac NL compounds include those displaying Kramers degeneracy \cite{rappe_dirac_line_nodes,C3P22,CaAsX}, those that do not possess it \cite{HgCr2Se4,hasan_TlTaSe2}, and even elements without mirror symmetry \cite{fcc}.

Recently, it was theoretically proposed that certain compounds might host a Dirac NL protected by the glide-mirror symmetry, a type of non-symmorphic symmetry, in their crystallographic structure \cite{Xu}. These compounds (ZrSi\emph{X}, \emph{X} = S, Se and Te) possess the ZrSiS-type structure \cite{Xu} (space group $P4/nmm$).
In agreement with the band-structure calculations \cite{Xu}, angle resolved photoemission spectroscopy (ARPES) studies on all ZrSi\emph{X} compounds \cite{schoop,neupane,ZrSiX_ARPES,ZrSiTe_ARPES} observed the predicted non-symmetry-protected band crossings at the $X-$points of their first Brillouin zones, in addition to symmorphic band crossings between the $ \Gamma-X $ and the $ M-\Gamma $ points.

Both ZrSiS and ZrSiSe display anomalous transport properties, such as the so-called ``butterfly" magnetoresistivity \cite{ali2,mao1}, which presents a maximum when the electrical current and the external magnetic field form an angle $\theta \simeq 45^{\circ}$, which does not correspond to a Lorentz force maximum. The relation between this butterfly MR and the NL remains unclear, although Ref. \onlinecite{ali2} claims to observe a topological phase transition upon rotation of the field. In contrast, ZrSiTe exhibits a two-fold-symmetric angular MR, which can be understood as resulting from the quasi-two-dimensional character of its FSs.

Here, we report electrical transport and magnetic torque measurements in both ZrSiSe and ZrSiTe with the goal of correlating their transport properties with the
experimentally determined electronic structure at the Fermi level. The dHvA effect reveals Fermi surfaces with geometries in reasonable agreement with band structure calculations which place the nodal lines at symmorphic and non-symmorphic positions in close proximity to the Fermi level $\varepsilon_F$ in ZrSiSe and in ZrSiTe, respectively.
It turns out that the transport lifetime in ZrSiSe is considerably longer than its quantum one and only this compound displays a four-fold symmetric magnetoresistivity as a function of field orientation. Given the acceptable agreement between the DFT calculations and our experimental results, and given their non-topological character according to our calculations, we conclude that the butterfly magnetoresistivity
bears little to no relation with the existence of nodal lines in these systems. Instead, and given the isotropic effective masses extracted for ZrSiSe, we find that this behavior correlates with its quasiparticle lifetime(s) as evaluated through its Dingle temperature(s). The Fermi surface geometry is remarkably different between both compounds despite their isostructural crystallinity with, most likely, this difference being at the origin of their contrasting magnetotransport behavior.

Information concerning sample synthesis and their characterization, single-crystal X-ray diffraction, and experimental setups are provided below in the Materials and Methods section.

\section{Experimental results: comparison with first-principles calculations}

Figures 1(A) and 1(B) display, respectively, the calculated FSs of ZrSiSe and of ZrSiTe within their first
Brillouin zone (FBZ), as obtained from DFT without spin-orbit coupling (WIEN2K + GGA, see Supplementary Materials section S2 for details of methodology). The FS of ZrSiSe is composed of nearly touching, but disconnected, electron-like pieces (in brown) around the $Z-$point composing a ``diamond" like surface with corrugation in the $k_x-k_y$ plane, akin to the one reported for ZrSiS \cite{schoop}.
Similarly to ZrSiS, the hole-like FSs in ZrSiSe (in violet) are three-dimensional (3D) in character, with the biggest pocket
located in the neighborhood of the $R-$point \cite{schoop}. In contrast, the FS of ZrSiTe presents a quasi-two-dimensional character displaying corrugated, electron-like cylindrical sheets (red color) along the $X-$ to $R-$direction, in addition to 3D pockets. The hole sheets (in green) are also corrugated cylinders, but of oblong cross-section and whose axis present a
sinusoidal modulation along the inter-planar direction. ZrSiTe also contains a second thin columnar hole surface, shown in blue.
Figures 1(C) and 1(D) show the dispersion of the relevant hole and electron bands throughout the entire $\Gamma-X-M$ plane for ZrSiSe and ZrSiTe, respectively, within their FBZ.  Without the SOC, both compounds show a $C_{2v}$ symmetry-protected Dirac nodal line (DNL) near the Fermi energy (shown as a blue plane).  According to our calculations, for both compounds, the Dirac-like
dispersion of the conduction and valence bands extends a few hundreds of meV away from $\varepsilon_F$, (see Fig. S1). The DNL is never more than 60 meV away from $\varepsilon_F$ for ZrSiSe and always in the positive energy direction, yielding only tiny hole pockets in this plane. In contrast, the DNL in ZrSiTe has a total disperson of over 400 meV, with $\varepsilon_F$ cutting near the energy midpoint, giving rise to sizable pockets of both hole and electron character, in agreement with the ARPES study in Ref. \onlinecite{hosen} which finds somewhat similar FSs for all three ZrSi\emph{X} compounds. This distance of the nodal lines with respect to $\varepsilon_F$ implies that its bulk charge carriers are unlikely to display anomalous transport properties from its topology.
Along with the Fermi surfaces in Figure 1(A) and 1(B), we plot the points in the $\Gamma-X-M$ plane where the non-SOC DNL falls within 80 meV of $\varepsilon_F$. For ZrSiSe, this results in several line segments, very close to or intersecting the Fermi level (black markers in Fig. 1(A)), which are interrupted by tiny islands of hole-like Fermi surface.  For ZrSiTe, only one point meets this qualification and falls just at
the edge of the large electron pockets.

Since the symmorphic nodal line (protected by $ C_{2\nu} $ symmetry) which is positioned at the center of the FBZ as depicted in the Fig. 1, is not protected against SOC so it completely gaps out. Therefore, conduction and valence bands remain separate throughout the entire FBZ. In this sense, ZrSiSe and ZrSiTe are topologically equivalents of an insulator \cite{Xu, TaPbSe2_PRB}. Given the presence of inversion symmetry, a $Z_2$-invariant could be acquired through the calculation of their Pfaffian \cite{Z2_Pfaffian}. Our calculations indicate that the SOC would gap the symmorphic NL by $\sim 36$ meV and $\sim 98$ meV in ZrSiSe and in ZrSiTe, respectively.
This has no discernable effect on the Fermi surfaces of either compound, i.e. it changes the calculated dHvA orbits by at most $\pm$ 5 T. For ZrSiTe, this is because the SOC-induced gap opens away from the Fermi energy and throughout the entire BZ (except at the single point indicated in Fig 1(B)), and the FSs that are formed away from these crossings do not experience any SOC shift. Given that the DNL in ZrSiTe is generally far from $\varepsilon_F$, the SOC does not gap any significant portion of its Fermi surface.

For ZrSiSe, the situation is somewhat different: since the symmorphic NL of ZrSiSe falls within a distance in energy that is comparable to the SOC-induced gap between bands, this system could be expected to gap in a way that is topologically equivalent to a topological insulator. Indeed, a finite gap opens at $\varepsilon_F$ where the valence and conduction bands previously touched forming the NLs. This gap does not reduce the original FS weight calculated without SOC because it already is zero at these band crossings even in the absence of a gap (inset of Fig. S1(a)). Aside from the carriers associated with the linearly crossing bands, ZrSiSe contains a region of ``normal" semimetallic character, seen predominantly along the $Z-R$ symmetry of the BZ where the valence and the conduction bands overlap in energy by $\simeq 300$ meV. These bands are separated in $k$-space, have a parabolic dispersion, and have no topological properties. SOC in this region has no effect within the computational accuracy and hence the semimetallic overlap is precisely maintained. This suggests that, at finite temperatures, one could expect to have a small portion of massive Dirac dispersive carriers
in the SOC-split NL bands, appearing in the $\Gamma-X-M$ plane, and some carriers of non-topological character in the semimetallic bands elsewhere in its BZ.

Figure 2 exhibits the magneto-transport properties of both compounds. ZrSiSe displays a very small residual resistivity $\rho_0$ with values
ranging from $\sim 50$ to $\sim 100$ n$\Omega$cm depending on the crystal and on its air exposure, which tends to degrade or oxidize the samples.
These values are close to those displayed by the best Cd$_3$As$_2$ single-crystals, a compound claimed to host an novel protection mechanism suppressing backscattering \cite{ong_Cd3As2}. Figure 2(A) displays the resistivity $\rho$ of a ZrSiSe single-crystal, for currents flowing along its $ b- $axis, as a function of the magnetic field $\mu_0H$ (up to 60 T) applied along several angles $\theta$ relative to its $ c- $axis. Here, we rotate the field within the $ac-$plane while always maintaining the current perpendicular to the field and along the $b-$axis, which is the condition that maximizes the Lorentz force. Remarkably, the magnetoresistivity displays a maximum for $\theta \simeq 50^{\circ}$ and not for fields applied perpendicularly to the conducting planes.
Notice that one observes the same angular pattern when the field is rotated from a direction parallel to the current towards a direction perpendicular to it (Fig. 2(E)). As seen in the inset for this field orientation, the resistivity increases by a factor $\geq 10^{6}$ when $\mu_0H$ reaches $\simeq 60$ T at $T = 4.2$ K, which is comparable to, or exceeds values observed in both WTe$_2$ \cite{ali} and the Weyl-type II semimetallic candidate WP$_2$ \cite{rico} at lower $T$s.
As shown in Fig. 2(B), for a second single-crystal with a relatively higher residual resistivity that exceeds 100 n$\Omega $cm, one still observes a very pronounced
increase $\rho(T)$ (exceeding $10^3$ at 9 T) below a characteristic temperature $T_{\text{min}}$. The inset shows $T_{\text{min}}$ as a function of $\mu_0H (T)$. The red
line is a linear fit. In contrast, ZrSiTe displays a far more pronounced $\rho_0$, i.e. $\gtrsim 20$ $\mu \Omega$cm, which is comparable to values reported by Ref.
\onlinecite{hosen} (see Fig. 2(C)). Its magnetoresistivity is also far less pronounced, with $\rho(\mu_0H)$ increasing by $\gtrsim 10^3$ \% when $\mu_0H$ reaches 31 T at
$T = 0.35$ K. In contrast to ZrSiSe, the maximum of $\rho(\mu_0H)$ in ZrSiTe is observed for fields nearly along the $ c- $axis. A better comparison between
both compounds is offered by the angular dependence of their magnetoresistivity which is displayed in Figs. 2(D) and 2(E) in polar coordinates, for ZrSiSe and ZrSiTe,
respectively. For ZrSiSe $\rho(\theta,\mu_0H)$ is predominantly four-fold symmetric \cite{ali2,mao1}. In contrast, ZrSiTe displays a conventional two-fold symmetric
magnetoresistivity \cite{mao1}, although an additional structure is observed at the highest field which is probably ascribable to the Shubnikov-de Haas effect.

In summary, Figs. 1 and 2 indicate that ZrSiSe displays very low residual resistivity, very pronounced magnetoresistivity, and high carrier mobilities (as discussed below)
which, at first glance, would seem to suggest that its bulk carriers display topologically non-trivial character. However, as discussed above, this interpretation is
at odds with our calculations pointing to trivial band topology in this compound. In contrast, ZrSiTe shows a high residual resistivity with a concomitantly smaller two-fold magnetoresistivity although it should be a better candidate for displaying anomalous transport properties due to the proximity of its non-symmorphic NL with respect to $\varepsilon_F$.

Our observations and calculations lead us to conclude that the butterfly magnetoresistivity bears no relation to the existence of the nodal lines or to their proximity with respect to the Fermi level, in contrast to what is claimed by Ref. \onlinecite{ali2}.
Hence, two- or four-fold symmetric angular magnetoresistivity as observed in either compound ought to result from the anisotropy of their Fermi surfaces, associated effective masses and quasiparticle lifetimes.

To address this point, we measured the geometry of the Fermi surfaces of both compounds via the angular dependence of the de Haas-van Alphen-effect (dHvA). Our goal is to compare our measurements with the calculations since a good agreement between both would confirm the pertinence and the concomitant implications of the calculations. Subsequently, We now show that the Fermi surface found from DFT is in relatively good agreement with the angular dependence of the dHvA effect. Figure 3(A) provides typical magnetic torque $\overrightarrow{\tau} = \overrightarrow{M} \times \mu_0 \overrightarrow{H} $ data as a function of the magnetic field and for two field orientations, respectively, along the $ c- $axis (top panel) and along the $ a- $axis (bottom panel). By subtracting the background signal through a fit to a polynomial, one obtains the dHvA signal which is plotted as a function of inverse field in Fig. 3(B). The oscillatory component in the magnetic torque $\Delta (\tau /\mu_0H)$ as a function of $(\mu_0 H)^{-1}$ can be described by the Lifshitz-Kosevich (LK) formalism \cite{Shoenberg}:

\begin{eqnarray}
 \begin{split}
  &\Delta (\tau /\mu_0H ) \propto  \\
  &- \frac{T}{(\mu_0H)^{3/2}}\sum_{i=1, l=1}^{\infty}  \frac{m^{\star}_i \exp^{-\pi l/ \omega_{c_i} \tau_i}\cos(l g_i m^{\star}_i \pi/2)}{l^{3/2}\sinh(\alpha m^{\star}_i T/(\mu_0H))}\\
  &\times \sin \left[2\pi \left( \frac{l \times F_i}{B} + \phi_{il} \right)\right]
 \end{split}
\end{eqnarray}

where $F_i$ is a dHvA frequency, $l$ is the harmonic index, $\omega_c$ the cyclotron frequency, $g_i$ the Land\'{e} \emph{g}-factor, $m^{\star}_i$ the effective mass in units of
the free electron mass $m_0$, and $\alpha = 14.69$ is a constant. Finally, $ \phi_{il} = (-\gamma_{il} + \delta_{il}$) is the Onsager phase,
where $\gamma = 1/2$ for a parabolic band or $ \gamma + \phi_B / 2\pi = 0 $ for a linearly dispersing one with $\phi_B$ being the Berry-phase.  $\delta$ is determined by the dimensionality of the FS taking values $\delta=\pm 1/8$ for minima and maxima cross-sectional areas of a three-dimensional FS, respectively.
Each dHvA frequency $F$ is related to an extremal cross-sectional areas $A$ of the FS through the Onsager relation $F = \hbar A/2 \pi e$. Hence, a Fourier transform of the oscillatory signal(s), as shown in Fig. 3(C), yields the fundamental $F$s whose dependence on field orientation can be directly compared with the FS extremal cross-sectional areas
determined by DFT in Fig. 1. This is done in Fig. 3(D) where we plot the Fourier spectra collected under several field orientations as a function of $F$. These traces were vertically displaced by the value of the angle $\theta$ between $\mu_0H$ and the $ c- $axis, at which each trace was collected. On this plot we have superimposed the angular dependence of the FS extremal cross-sectional areas according to DFT (converted into $F$s \emph{via} the Onsager relation) for the electron-pockets that form the diamond-like FS (see Fig. 1), for the larger 3D hole-pockets, and for the ring of tiny 3D hold pockets. As shown, the angular dependence of the hole pockets is well described by the calculations. As for the electron pockets, we observe multiple peaks as the field is rotated away from the $ c- $axis, which probably results from their corrugation (maximal and minimal cross-sections) and from an imperfect alignment of $\mu_0H$ with the $ a- $axis. Their angular dependence is also captured by the calculations.
In another crystal and for fields rotating within the planes we also observe a good agreement between the DFT and the measured dHvA frequencies (Fig. S2). The effective masses associated with most of the fundamental frequencies display values in the order of 0.1-0.2 $ m_0 $, again in agreement with DFT (Fig. S3).

It is therefore pertinent to ask if one can extract a non-trivial $ \phi_B =\pi$ for any of the observed orbits.
To this effect, we fit the oscillatory components of the dHvA signal of ZrSiSe, collected from both SQUID and torque magnetometry, using the Lifshitz-Kosevich (LK) formalism (Eq. 1). It turns out that the smaller frequencies are found to be field-dependent (Fig. S4), while the large number of frequencies and associated fitting parameters makes it nearly impossible to extract reliable, or unique values, for $\phi_B$. For the few orbits where we managed to evaluate $\phi_B$, we ended up obtaining \emph{trivial} values.

Figure 4(A) displays two representative $\tau$ traces for a ZrSiTe single-crystal as a function of $\mu_0H$ applied nearly along the $c-$ and the $a-$axis, respectively. The superimposed oscillatory or dHvA signal is shown in Fig. 4(B) as a function of $(\mu_0H)^{-1}$ while the respective Fourier transforms are shown in Fig. 4(C). The FFT spectra display less frequencies than the ones extracted from ZrSiSe, which is partially due to the absence of the frequencies associated with the small hole-pockets in the $\Gamma-X-M$ plane. In Fig. 4(D) we superimpose the angular dependence of the calculated dHvA frequencies (dashed lines) onto the
experimentally observed ones. Dashed lines in blue and in red colors correspond to calculated frequencies associated with the electron-like FS sheets, while green dashed line corresponds to the frequency of the smallest hole-orbit.
As shown, the position of the calculated and of the experimentally observed electron orbits agree well,
although a better agreement would be obtained through a very small displacement of $\varepsilon_F$.
The dHvA signal extracted from ZrSiTe via torque and SQUID magnetometry techniques for fields nearly along the $c-$ and the $a-$axes was also fit to one or two LK-components yielding trivial values for the Berry-phase (Fig. S5). The effective masses associated with the electron pockets of ZrSiTe display values in the order of $\mu \simeq 0.1$ $m_0$ thus confirming that its electronic dispersion at the Fermi level is nearly linear (Fig. S6). The absence of a heavier pocket is consistent with the lack of an associated quadratic dispersion in the band structure.

\section{Discussion}

Our study on the Fermi surfaces of ZrSiSe and ZrSiTe \emph{via} the dHvA-effect reveals a general good agreement with the predictions from
Density Functional Theory calculations.
Despite being isostructural, ZrSiSe and ZrSiTe are characterized by very different FSs (see, Fig. 1). ZrSiSe displays three dimensional FSs and exhibit exotic magneto-transport properties, such as an extremely large MR, ultra-high carrier mobilities and an intriguing butterfly-like angular magnetoresistivity. These properties have been associated to the existence of the Dirac nodal lines and, tentatively, to the non-trivial topology \cite{ali2} of its electronic bands. However, our calculations indicate that the nodal lines located in the vicinity of $ \varepsilon_F $ in ZrSiSe are not stable with respect to gap opening in the presence of SOC.
The SOC-induced gap opening throughout the entire BZ converts this system into the topological equivalent of an insulator in the bulk (TI) whose topological character is dictated by a set of $Z_2$ invariants (i.e. $ \nu_0;(\nu_1\nu_2\nu_3) $) \cite{fu} with $ \nu_0 $ indicating a strong TI index.
The weak TI related $Z_2$ invariant obtained for ZrSiO, see Ref. \onlinecite{Xu}, implies that this family of isostructural compounds are likely to be characterized by trivial topology. Indeed, according to our calculations the $Z_2$ index ($ \nu_0 $) for ZrSiSe is zero. A vanishing $ \nu_0 $ is also found for other isostructural compounds, see Refs. \onlinecite{ZrSiS_Sinica,ZrSiX-PRX}, and this is consistent with Ref. \onlinecite{ZrSiX-PRX} whose DFT analysis provides a non-topological explanation for the observed surface states in ARPES.

With these compounds being topologically trivial, one cannot attribute the butterfly-like angular magnetoresistivity of ZrSiSe to the topological character of its bulk electronic band-structure. Note that this butterfly AMR emerges at fields as low as $ \mu_0H \sim $ 5kOe (Fig. S7), which corresponds to a rather small energy scale to induce a topological phase-transition.
Therefore, the origin of the butterfly AMR ought to result from a conventional mechanism. For example, according to the two-band model \cite{Ashcroft} which is based upon semi-classical Boltzmann transport theory, the magnitude of the magnetoresistivity is proportional to the average mobility of the charge carriers  ($ \mu_{\text{av}} $) in a compensated system, which is given by the effective masses $ m^{\star} $ and the quasiparticle scattering rate $ 1/\tau $, that is $ \mu = \frac{e \tau}{m^{\star}} $.
The descent agreement of angular dHvA frequencies with DFT predictions proves the carrier compensation in ZrSiSe and, hence, DFT-calculated electron and hole densities i.e. $ n_e \simeq 1.814 \times 10^{20}\ \textrm{and}\ n_h \simeq 1.887 \times 10^{20}\ \textrm{cm}^{-3} $, respectively, are implanted in our two-band and modified-two-band methods. In a compensated system ($ n_e \approx n_h $), the average carrier mobility can be expressed as $ \mu_{av} = \frac{2 \mu_e \mu_h}{\mu_e + \mu_h} $.
We extracted the electron and hole mobilities ($ \mu_e$ and $\mu_h $, respectively,) of ZrSiSe from fittings to the two-band and the modified-two-band models (See Fig. S7), as depicted in the Figs. 5(A) and (B), respectively.
The average mobility indeed displays the same butterfly-like angular dependence observed for the AMR. Hence, from this observation one would expect a strong angular dependence for the combined effective masses of ZrSiSe which could result from the anisotropy of its FSs \cite{Ashcroft} (see, Fig. 1(A)), leading to its angular-dependent mobility. However, our experimental determination of the combined effective masses, shown in Fig. 5(C), do not support the existence of a minimum around the angle where the maximal mobility is observed (that is $ \theta \sim 45-60^{\circ} $). Therefore, we conclude that these relatively isotropic effective masses seen in Fig. 5(C) indicate that the scattering rate ought to be dependent on field orientation. In order to confirm this hypothesis, we extracted the Dingle temperature as a function of the angle with respect to the external field through fittings of the raw dHvA signal to the LK formalism as discussed below. From our LK fittings to the experimental dHvA signal we detected a modulation on the Dingle-factor $T_D$ which displays values between 1.5 and 2.5 K at angles exceeding $\theta = 35^{\circ}$ (Figs. S4, S8, and Table S1 displaying the $T_D$ values extracted from the sample in Fig. S8). But $T_D$ increases to values of $\sim 6.5$ and $\sim 10$ K for the frequencies that one associates with the electron and the hole-pockets, respectively, when the field is aligned nearly along the $ c- $axis. This indicates that the sample behaves as if it had a higher scattering rate or a higher density of impurities when the field is aligned nearly along the $ c- $axis which suppresses its magnetoresistivity (increases the probability of carrier scattering before it can complete a cyclotron orbit) and produces a dip in its angular dependence. The effect of this modulation on the Dingle term can be directly judged from the Fourier transforms of the oscillatory signals shown in Fig. 3(D). In Fig. 3(D) the amplitude of the peak at the frequency that one associates with the electron pockets decreases considerably as the field is oriented towards the $c-$axis, becoming considerably smaller than the peaks observed at frequencies that one associates with magnetic-breakdown orbits among these electron pockets
(centered around 470 T). In the Shubnikov-de Haas data for another ZrSiSe crystal (see also, Fig. S2), one cannot even detect the electron orbits when $ \theta \leq 30^{\circ} $ (indicated by brown line). Hence, our data indicates that the electron pockets are particularly susceptible to the modulation of the Dingle temperature induced by the field. The dip observed for fields along a planar direction results from either the suppression of the Lorentz force (for the $\overrightarrow{H}\|\overrightarrow{j}$ configuration) or from inter-planar scattering (for the $\overrightarrow{H} \bot \overrightarrow{j}$ configuration) probably due to stacking faults or some level of stacking disorder.

Having exposed that the scattering rates associated with the electron and hole pockets in ZrSiSe display an angular dependence as the sample is rotated with respect to the external field, we now focus on the role of the field in inducing this anisotropy in the ZrSi\emph{X} compounds.
One possibility is that the $\mu_i$s and the $\tau_i$s, where $i$ is an index identifying a particular FS sheet, are field-dependent due to the action of the Zeeman-effect on small FS pockets. This effect could be analogous to the one reported for the Dirac semi-metal Cd$_3$As$_2$ where remarkably small, but sample-dependent, residual resistivities were observed and attributed to a novel protection mechanism that suppresses carrier backscattering under zero magnetic-field. In Cd$_3$As$_2$ the transport lifetime was found to be $10^4$ times longer than the quasiparticle or quantum lifetime observed under field \cite{ong_Cd3As2}. We observe a similar effect in ZrSiSe albeit we do not attribute it to a protection mechanism, but tentatively to Fermi surface evolution under field. To expose this effect we draw a direct comparison with Cd$_3$As$_2$ by plotting in Fig. 6 the conductivity $\sigma_{xx}$ and the Hall-conductivity $\sigma_{xy}$ of ZrSiSe, obtained by inverting its resistivity tensor. As discussed within Ref. \onlinecite{ong_Cd3As2}, $\sigma_{xy}(\mu_0H)$ exhibits a characteristic ``dispersive-resonance'' profile displaying sharp peaks that reflect the elliptical cyclotron orbits executed by the carriers under weak fields. According to Ref. \onlinecite{ong_Cd3As2}, its reciprocal value $(\mu_0H_{\text{max}})^{-1}$ would, within standard Bloch-Boltzmann transport theory, yield the geometric mean value of the transport mobility $\overline{\mu_{\text{tr}}}$ within a plane perpendicular to the field. From the reciprocal of $\mu_0H_{\text{max}}$ we obtain $\overline{\mu_{\text{tr}}} \simeq 1.89 \times 10^5$ cm$^2$/Vs which is comparable to values quoted in Ref. \onlinecite{ong_Cd3As2},
although some of their Cd$_3$As$_2$ crystals displayed values in excess of $1 \times 10^6$ cm$^2$/Vs. Hence, we obtain a mean transport lifetime $\overline{\tau_{\text{tr}}} = \overline{\mu_{\text{tr}}} \mu/e = 109.4 \times 10^{-12}$ s, where $\mu \simeq 0.1$ $m_0$ is the typical carrier effective mass for ZrSiSe, and $e$ is the electron charge. We have checked that this method yields decreasing values for $\overline{\tau_{\text{tr}}}$ as $T$ increases due to phonon scattering (Fig. S9). This $\overline{\tau_{\text{tr}}}$ value contrasts markedly with the quantum or quasiparticle lifetime as extracted from the Dingle damping factor: all fittings of the dHvA signal to the LK formalism shown in Figs. S4 and S5 yield Dingle factors $T_D$ ranging from 1 to 2 K for the best crystals to $T_D \simeq 10-20$ K for the lower quality ones. These imply quantum lifetimes $\tau_Q = (\hbar/2 \pi k_B T_D)$ ranging from $0.05 \times 10^{-12}$ s to $1 \times 10^{-12}$ s which are $\sim 10^3 $ to $\sim 10^2$ times smaller than $\overline{\tau_{\text{tr}}}$. Hence, we conclude that the field modifies the geometry of the smallest FS sheets via the Zeeman-effect leading to quantum lifetimes that are considerably shorter than the transport ones.

In contrast, for ZrSiTe we find similarly light effective masses $ \sim 0.1\ m_e $ but shorter transport lifetimes ($\overline{\tau_{\text{tr}}} =1.9 \times 10^{-12}$) with respect to those of the Se compound. An analysis of the Dingle-factor yields $ T_D\ \simeq\ 14$ K, which is in the same range as those of ZrSiSe when $ \mu_0H $ is close to the $c-$axis. Therefore, in ZrSiTe the quantum lifetime is comparable to the transport lifetime although, according to band structure calculations (see, Fig. S1), it displays Dirac nodal lines protected by non-symmorphic symmetry located relatively close to the Fermi level and therefore it should have been a better candidate for hosting non-symmorphic Dirac fermions.

In summary, we suggest that our observations present a certain analogy with Cd$_3$As$_2$ where the application of a field is claimed to reconfigure the Fermi Surface sheets, but in our case it affects the scattering processes (see Ref. \cite{ong_Cd3As2}) albeit in the absence of topological protection at zero field. This effect is anisotropic probably due to the spin-orbit coupling that is likely to lead to an anisotropic Land\'{e} \emph{g}-factor. Our results indicate that the field plays a more prominent role when it is applied perpendicularly to the plane of the diamond shaped Fermi-surface(s) becoming less effective when it is rotated towards its plane.

A final word about the evaluation of the topological character of ZrSiSe. We indeed observed negative longitudinal magnetoresistivity, claimed in Dirac and Weyl systems to result from the chiral anomaly among Weyl points \cite{ong}. But after a detailed analysis on the current distribution \cite{ong}, we concluded that it results from the so-called current jetting effect \cite{ong}, i.e. when voltage contacts are placed along the spine of the crystal, the magnetoresistivity always has positive slope w.r.t $ \mu_0H $ . The so-called planar Hall effect (PHE), proposed to be particularly pronounced in Weyl systems due to the chiral anomaly \cite{burkov}, is also clearly seen in this compound (Fig. S10). Nevertheless, by exfoliating
ZrSiSe down to about 100 nm in thickness, we observe a dramatic decrease in both the longitudinal magnetoresistivity and the PHE signal becomes barely observable (data not included). It is difficult to understand this observation in the frame of the axial anomaly that should remain oblivious to sample thickness. Instead, it suggests that that the PHE in ZrSiSe is merely a magnetoresistive effect, as seen, for instance, in elemental Bi \cite{ong}, associated with sample geometrical factors and their interplay with the carrier mean free path.

\section{Conclusions}
DFT calculations predict Fermi surfaces for ZrSi(Se,Te) which are in close agreement with our de Haas-van Alphen study, implying a modest role for the spin-orbit coupling and for
electronic correlations (under zero field). In the absence of spin-orbit coupling it also predicts for ZrSiSe, but not for ZrSiTe, a close proximity of its Dirac nodal line to the Fermi level. In absence of spin-orbit coupling sections of the nodal line would intersect the $\Gamma-X-M$ plane within the Brillouin zone of ZrSiSe, leading to Dirac like
quasiparticles under zero-field. Nevertheless, spin-orbit coupling is expected to entirely gap this Dirac nodal line. This is not the case for ZrSiTe which according to our calculations present symmetry protected nodal lines relatively close to the Fermi level.

We conclude that ZrSiSe is akin to an increasing number of materials predicted to host novel band degeneracies, for which there is a clear lack of strong experimental evidence supporting claims in favor of nontrivial topology. Take for example, Refs. \onlinecite{trivial-LaAs,trivial-LaSb,trivial-MoAs2,trivial-TaAs2} revealing intriguing transport properties for all of these compounds, namely LaAs, LaSb, MoAs$ _2 $ and TaAs$ _2 $ which turn out to be topologically trivial due to the spin-orbit coupling lifting the degeneracy between band crossings and thus leading, in this particular system, to a vanishingly small $Z_2$ invariants. Although, in our case, the glide mirror symmetry does protect the degeneracy of the nodal loops located far away from $ \varepsilon_F $, the SOC transforms ZrSiSe from a Dirac nodal-loop to a trivial semimetal by opening an energy gap along the $ C_{2\nu} $ nodal loop. According to our calculations this leads to a topologically trivial $Z_2$ invariant for ZrSiSe. Its very small residual resistivity and its large, anomalous angle-dependent magnetoresistivity cannot be associated with its nodal lines. In fact, we have shown here that the butterfly magnetoresistivity results from the field-induced modulation of carrier mobilities, or in fact quasiparticle lifetimes, imposed by field orientation. In contrast, ZrSiTe does not reveal any novel or unconventional transport property that one could associated with non-trivial band topology or with some particular effect of the external field.

Despite the trivial topological character of its electronic bands, ZrSiSe presents strong similarities with the Dirac semimetal Cd$_3 $As$_2$, such as a pronounced non-saturating magnetoresistivity, low residual resistivity, high mobilities (approaching $ 10^6 $ cm$^2$/Vs) and very light effective masses ($ \sim 0.1\ m_e$). In Cd$_3 $As$_2$ these properties were attributed to the nontrivial topological character of its electronic dispersion which, as we discussed above, is not the case for ZrSiSe. Given the similarities between both compounds, it is natural for us to ascribe the butterfly magnetoresistivity to a mechanism akin to the one proposed for Cd$_3$As$_2$, for which the magnetic field was suggested to affect the geometry of its Fermi surface and hence its related effective masses and quasiparticle lifetimes. But we did not detect any marked effect of the field, or of its relative orientation, on the effective masses of the quasiparticles. Instead, our results indicate, quite clearly, that the field affects the carrier lifetimes with this effect being anisotropic or more pronounced for fields along the inter-planar direction. Hence, this leads to a four-fold symmetric butterfly-like magnetoresistivity as a function of the angle between the field and any given crystallographic axis.

\section{Materials and Methods}

High quality ZrSiSe crystals were synthesized by a solid state reaction using a 1:1:1 molar ratio of Zr, Si, and Se were directly reacted in vacuum sealed quartz ampoules at 950 $^{\circ}$C for 21 days. Plate-like single-crystals with average dimensions of $ \sim 300 \times 300 \times 150$ cm$^3$ were obtained. An $I_2$ mediated chemical vapor transport technique was also employed for the synthesis of ZrSi(Se,Te). The obtained ZrSiSe crystals were post-annealed for a week at 650 $^{\circ}$C. Annealing was not applied to ZrSiTe given its chemical instability at high temperatures. As discussed below, single-crystal X-ray diffraction measurements confirm that both ZrSiSe and ZrSiTe crystallize in the space group $P4/nmm$. Magnetization measurements at low fields were performed in a commercial superconducting quantum interference device (SQUID) magnetometer. Magneto-transport property measurements were performed through a standard four-terminal configuration in multiple magnets, including 35 T and 31 T Bitter resistive magnets coupled to either 3He refrigerators or variable temperature inserts at the National High Magnetic Field Laboratory in Tallahassee. Cantilever beam torque magnetometry measurements were performed simultaneously with the transport measurements performed in continuous fields via a capacitive and piezoresistivity method. Pulsed-fields up to 62 T with duration of 150 ms were provided by the Dresden High Magnetic Field Laboratory. 
The magneto-transport measurements have conventional four-terminal configuration and, except the Fig. 2(D), the electrical current is applied along the crystallographic $ b- $axis while the magnetic field is rotating in the $ ac- $plane.

\section{supplementary materials}
\noindent
Section S1. Structural determination through X-ray\\
Section S2. Band Structure calculations\\
Fig. S1 Density Functional Theory electronic band structures for ZrSiSe and ZrSiTe\\
Fig. S2. Fast Fourier transforms of the dHvA signal of ZrSiSe as a function of field orientation\\
Fig. S3. Fourier transform of the dHvA signal of ZrZiSe under several temperatures\\
Fig. S4. Raw dHvA signal of ZrZiSe as a function of inverse field $ (\mu_0H)^{-1} $ and fits to the LK formalism\\
Fig. S5. Raw dHvA signal of ZrZiTe as a function of inverse field $ (\mu_0H)^{-1} $ and fits to the LK formalism\\
Fig. S6. Fourier transform of the dHvA signal of ZrZiTe under several temperatures\\
Fig. S7. Magnetoresistance of ZrSiSe and fittings to the two band model\\
Fig. S8 Fittings of the dHvA signal of ZrSiSe to the LK formalism\\
Table S1. Parameters extracted for ZrSiSe from fits to the multicomponent Lifshitz-Kosevich formalism\\
Fig. S9. Estimation of the carrier mobility of ZrSiSe at $ T $ = 30 K\\
Fig. S10. Planar Hall effect in ZrSiSe\\

\section{References}

\section{Acknowledgments}
\begin{acknowledgments}
	
	This work was supported by DOE-BES through award DE-SC0002613.
	K.-W.C. was partly supported by NHMFL-UCGP. The NHMFL is supported by NSF through NSF-DMR-1157490 and the
	State of Florida. We acknowledge the support of the HLD-HZDR, member of the European Magnetic Field Laboratory (EMFL).
	S. S. and K. Y. are supported by NSF-DMR-1442366
	V. L. Q. is supported by NSF-DMR-1555163
	
\end{acknowledgments}

\section{Figures}
\begin{figure*}[htb]
	\begin{center}
		\includegraphics[width=17 cm]{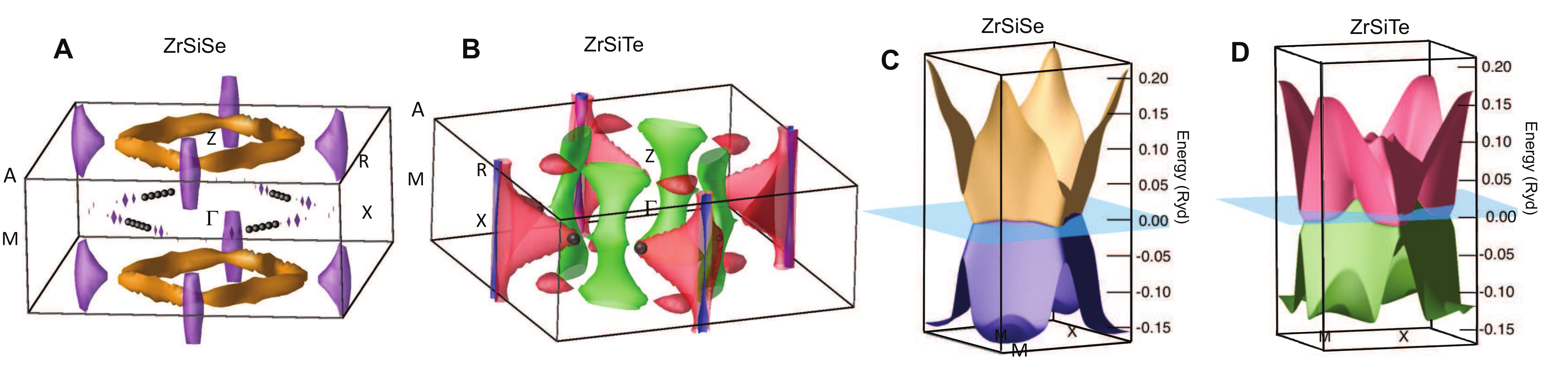}
		\caption{\textbf{DFT Fermi surfaces and band dispersions around $k_z =0$ without SOC.}
			(\textbf{A}) Calculated Fermi surfaces of ZrSiSe as obtained \emph{via} Density Functional Theory. Brown sheets are electron-like while violet ones are hole-like Fermi surfaces.
			(\textbf{B}) Calculated Fermi surfaces of ZrSiTe, where red colored sheets are electron-like and green ones are hole-like Fermi surfaces. (\textbf{C}) Energy $\varepsilon$ dispersion as a function
			of $k-$vector for ZrSiSe. Notice the linear dependence of $\varepsilon$ on $k$ for both the conduction (beige) and the valence bands (violet). Both bands meet at a
			nodal Dirac-line (line of Dirac nodes) close to the Fermi level $\varepsilon_F$ (clear blue plane). (\textbf{D}) Energy-dispersion bands for ZrSiTe showing that the dispersive nodal-line sits either above or below the Fermi-level throughout the Brillouin zone. In (A) the dots within the $X-M$ plane indicate the position of the Dirac nodes, composing the Dirac nodal line of ZrSiSe, which intersect $\varepsilon_F$. In ZrSiTe the Dirac nodes intersect just the tips of the electron-like (red colored) FSs extending along the $X-M$ plane. }
	\end{center}
\end{figure*}

\begin{figure*}[htb]
	\begin{center}
		\includegraphics[width=17 cm]{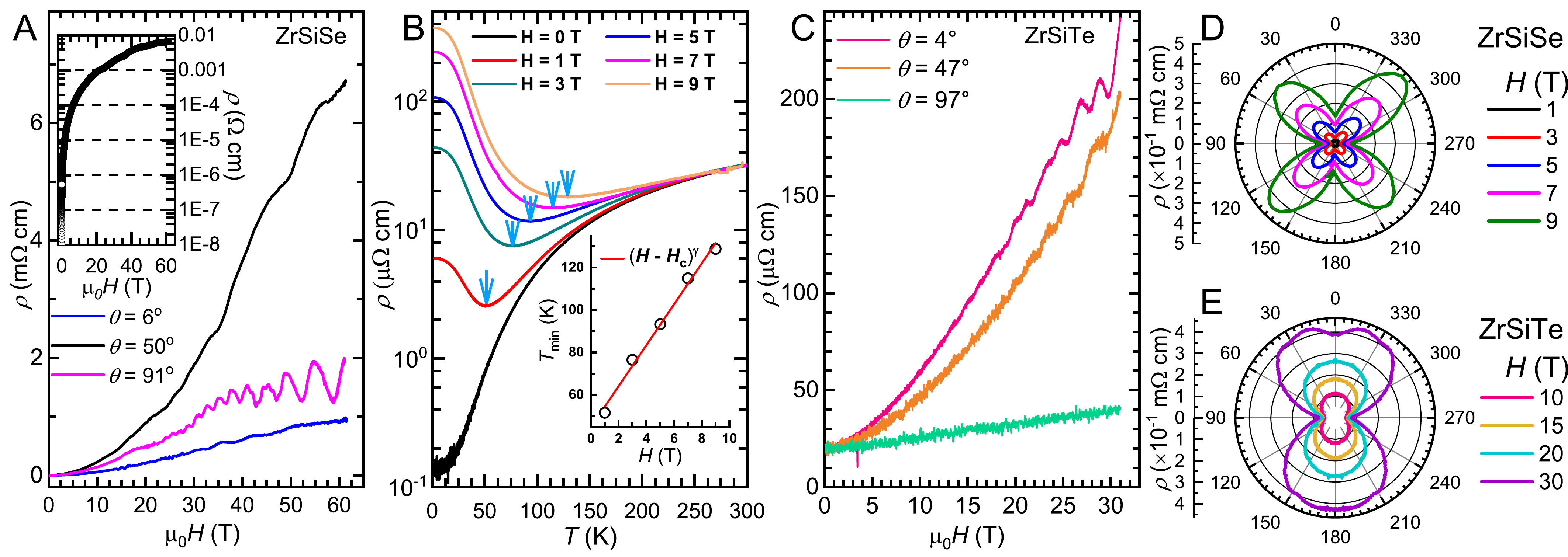}
		\caption{\textbf{Magneto-transport in ZrSiSe and ZrSiTe.}
			(\textbf{A}) Resistivity $\rho$ as a function of the magnetic field $\mu_0H$ for a ZrSiSe single-crystal. These traces were acquired at $T = 4.2$ K and for several angles $\theta$ between $\mu_0H$ and the crystalline $c-$axis. Inset: semi-log plot of $\rho (\theta = 45^{\circ})$ as a function of $\mu_0H$ indicating that $(\rho (\mu_0H)- \rho_0)/\rho_0$ reaches $\sim 10^8$  \% for this orientation. $\rho_0$ is the value of the resistivity at zero-field. (\textbf{B}) $\rho$ as a function of $T$ for a ZrSiSe crystal under several values of $\mu_0H\| c-$axis. Blue arrows indicate the position of the minima in $\rho(T,\mu_0H)$. Inset: position of the resistive minima, or $T_{\text{min}}$, as a function of $\mu_0H$. For these traces the electrical current was injected along the $ab-$plane. (\textbf{C}) $\rho$ as a function of $\mu_0H$ for a ZrSiTe crystal at different angles $ \theta$ under $T = 0.35$ K. $ \theta $ is the angle between $\mu_0H$ and the $c-$axis. (\textbf{D}) and (\textbf{E}) Angular-dependence of the magnetoresistivity for ZrSiSe and ZrSiTe, respectively. For all fields, ZrSiSe displays a ``butterfly" shaped magnetoresistivity as a function of  $\theta$. $\rho(\theta)$ is two-fold symmetric for ZrSiTe. However,  higher periodicities begin to emerge as $ \mu_0 H$ approaches or exceeds 30 T.}
	\end{center}
\end{figure*}

\begin{figure*}[htb]
	\begin{center}
		\includegraphics[width=13 cm]{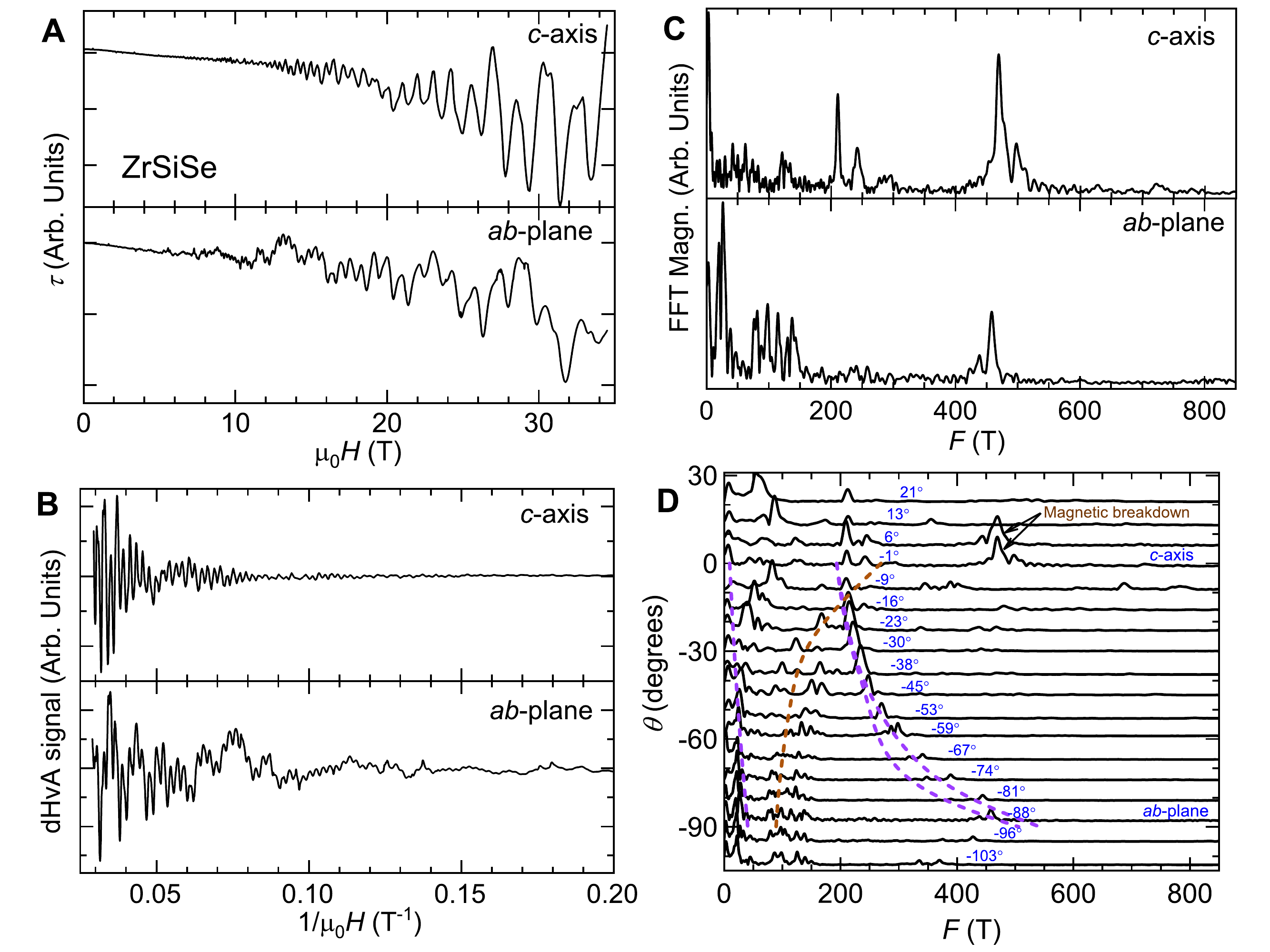}
		\caption{\textbf{Angular dependence of the dHvA-effect in ZrSiSe and its comparison with DFT calculations.}
			(\textbf{A}) Upper panel: Magnetic torque $\tau$ as a function of $\mu_0 H \parallel c-$axis for a ZrSiSe single-crystal at $T \simeq 1.4$ K. Lower panel: $\tau$ as a function of $\mu_0 H \parallel ab-$plane for $T \simeq 1.4$ K. (\textbf{B}) Superimposed oscillatory or de Haas-van Alphen signals in (A) as functions of $(\mu_0H)^{-1}$ obtained by subtracting the background torque \emph{via} a polynomial fit. (\textbf{C}) Corresponding fast Fourier transforms (FFTs) of the oscillatory signals. (\textbf{D}) FFTs as functions of the frequency $F$ for several angles $\theta$ between the external field and the $c-$axis. Dashed lines depict the angular dependence of the frequencies, or Fermi surface cross-sectional areas, according to the DFT calculations. Here, we have followed the same color scheme previously used to depict the Fermi surfaces of ZrSiSe in Fig. 1(A) to indicate orbits on the hole- or on the electron-like FSs, respectively.}
	\end{center}
\end{figure*}

\begin{figure*}[ptb]
	\begin{center}
		\includegraphics[width=13 cm]{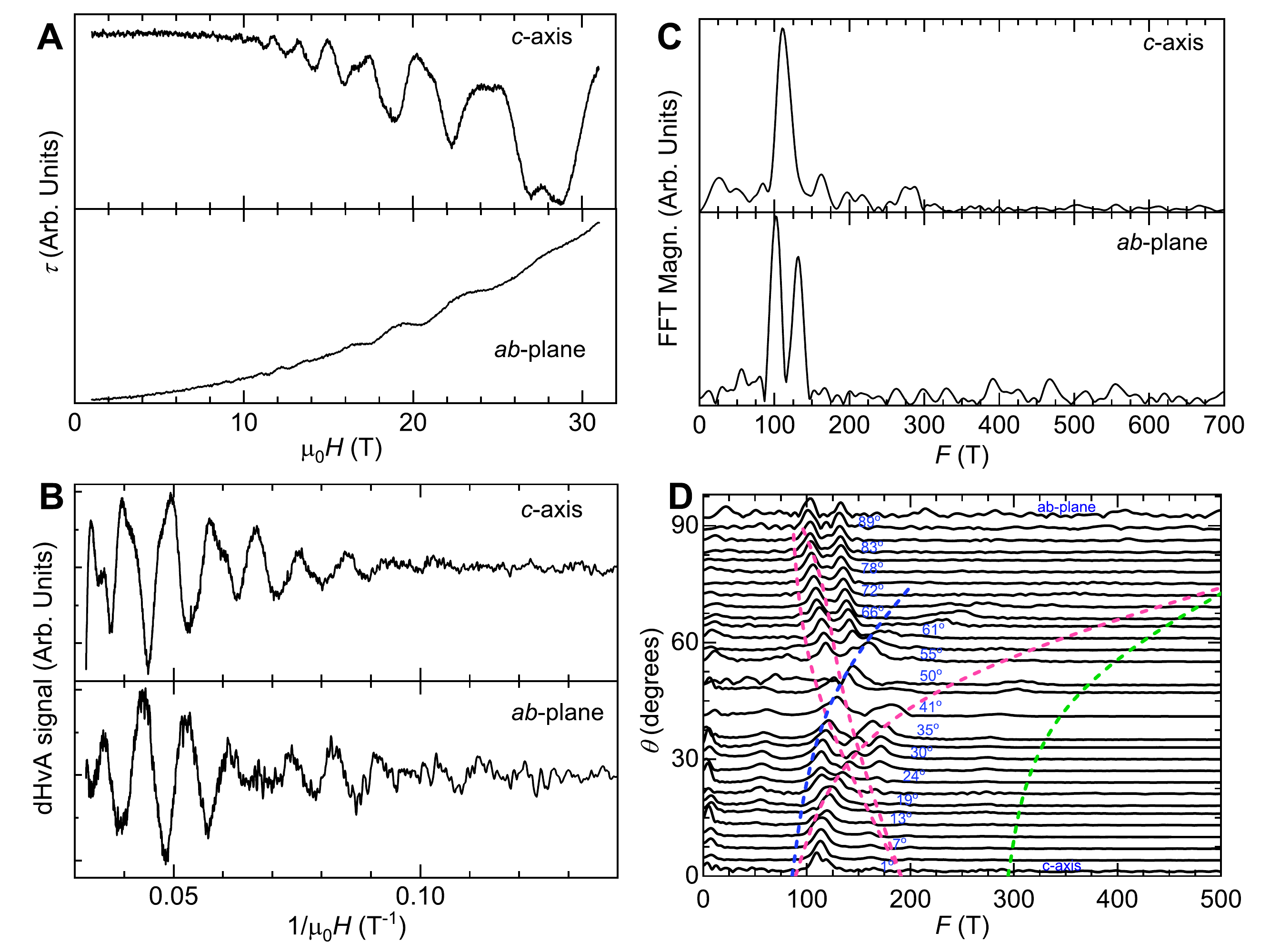}
		\caption{\textbf{Angular dependence of the dHvA-effect in ZrSiTe and its comparison with DFT calculations.}
			(\textbf{A}) Upper panel: Magnetic torque $\tau$ as a function of $\mu_0 H \parallel c-$axis for a ZrSiTe single-crystal at $T \simeq 0.4$ K. Lower panel: $\tau$ as a function of $\mu_0 H \parallel ab-$plane at the same $T$. (\textbf{B}) Superimposed dHvA signals in (A) as functions of $(\mu_0H)^{-1}$. (\textbf{C}) Corresponding FFTs of the oscillatory signals. (\textbf{D}) FFTs as functions of the frequency $F$ for several angles $\theta$ between the $\mu_0H$ and the $c-$axis. Dashed lines depict the angular dependence of the frequencies predicted by the DFT calculations. Again, their color indicate orbits on the hole- or on the electron-like Fermi surfaces.}
	\end{center}
\end{figure*}

\begin{figure*}[htb]
	\begin{center}
		\includegraphics[width=17 cm]{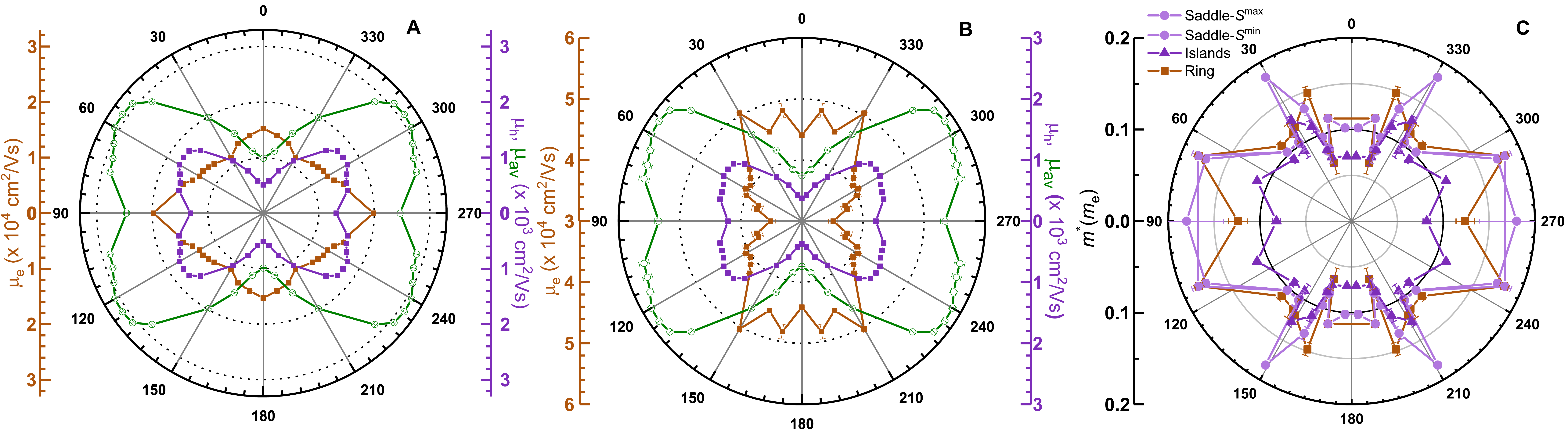}
		\caption{\textbf{Carrier mobilities and effective masses of ZrSiSe as a function of the angle.}
			(\textbf{A}) Electron, hole and average mobilities ($ \mu_e,\ \mu_h\ \textrm{and}\ \mu_{_{\textrm{av}}} $, respectively) extracted through fittings of the magnetoresistivity to the two-band model within the range of magnetic fields $ \mu_0H\ =\ 3-9 $ T. (\textbf{B}) The same data in (A) but resulting from a fit to a modified-two-band model including a linear term (see details in Supplemental Material, Fig. S7 \cite{supplemental}). In Ref. \onlinecite{ZrSiS_Sinica}, a linear in field component for the MR is observed for ZrSiS. Here, our modified-two-band model includes also a linear $ \mu_0H $ term.
			Despite the different fitting models, both panels (A) and (B) do display a butterfly-like angular dependence for their average mobilities. (\textbf{C}) Effective masses obtained by fitting the LK-formalism to the dHvA signals collected at different field orientations and whose frequencies can be tracked through the whole angular range. Saddle-$S^{\text{max}}$ and saddle-$S^{\text{min}}$ correspond to the same hole-pocket and represent its maximal and minimal cross-sectional areas, respectively. The color code is consistent with the one used to depict the Fermi Surfaces in Fig. 1(A).}
	\end{center}
\end{figure*}

\begin{figure}[htb]
	\begin{center}
		\includegraphics[width=8.6 cm]{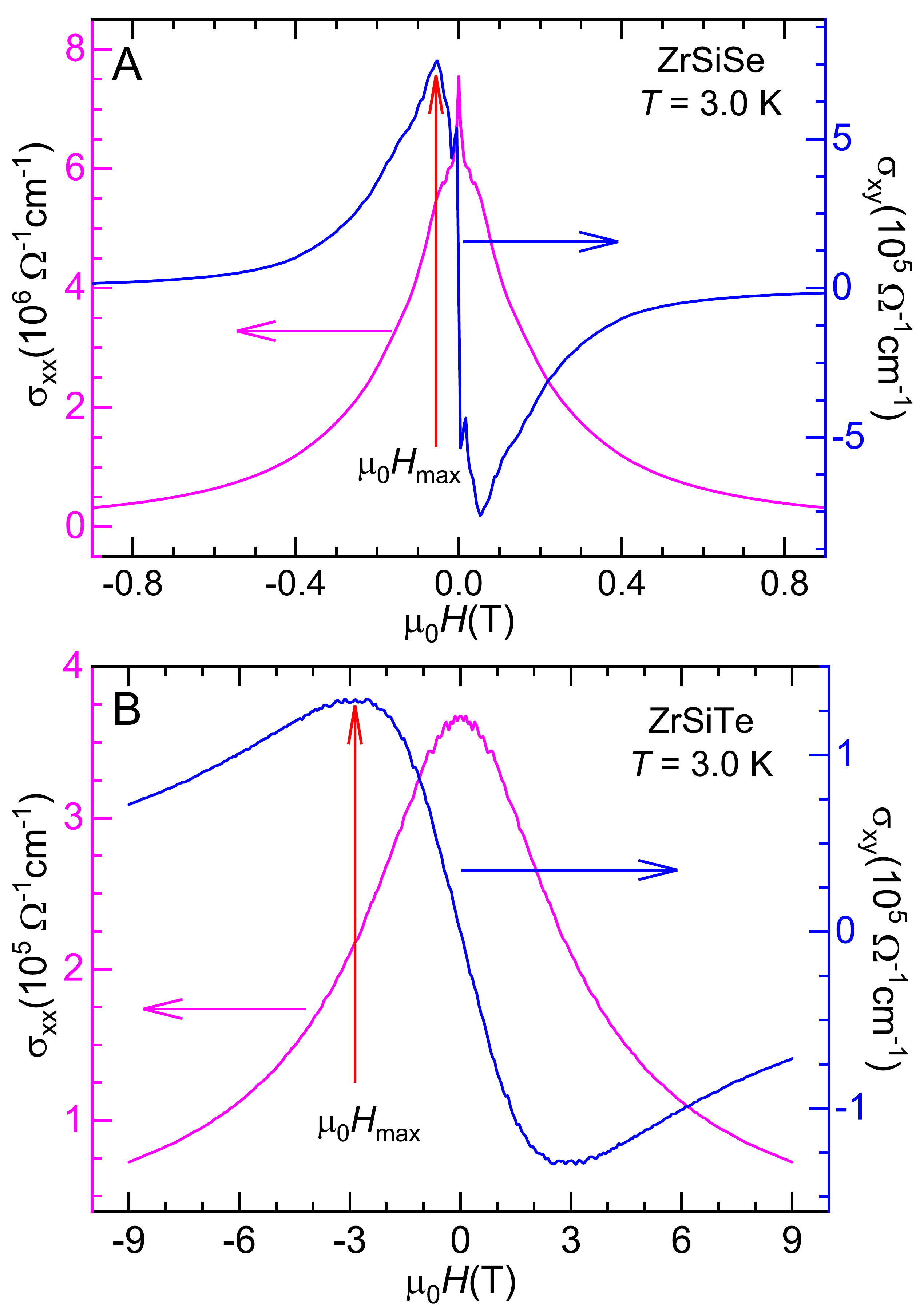}
		\caption{\textbf{Conductivity and Hall conductivity as a function of the field $\mu_0H$ for ZrSiSe and ZrSiTe.}
			(\textbf{A}) and (\textbf{B}) Conductivity $\sigma_{xx}$, and Hall conductivity $\sigma_{xy}$, for a ZrSiSe and ZrSiTe single-crystal, respectively, at $T = 3.0$ K. These were obtained by inverting the components $\rho_{ij}$ of the resistivity tensor as functions of $\mu_0H \|c-$axis. $\sigma_{xy}(\mu_0H)$ exhibits the characteristic ``dispersive-resonance" profile discussed in detail within Ref. \onlinecite{ong_Cd3As2} which displays a sharp peak at $\mu_0H_{\text{max}}$ reflecting the elliptical cyclotron orbit executed at low fields. From the position of the peak, i.e. $\sim 53$ \& $ 3000 $ mT, one extracts a mean transport mobility $\overline{\mu_{\text{tr}}} = (\mu_0H_{\text{max}})^{-1} \simeq 1.14 \times 10^5$ \& $ 3.34 \times 10^3 $ cm$^2$/Vs for ZrSiSe and ZrSiTe, respectively. }
	\end{center}
\end{figure}

\end{document}